\documentclass[conference]{IEEEtran}
 
\ifCLASSINFOpdf
\else
\fi

\usepackage{amsmath}
\usepackage[pdfborder={0 0 0}]{hyperref} 
\usepackage{dblfloatfix}
\usepackage{svg}
\usepackage{caption} 
\usepackage[nolist]{acronym} 
\begin{acronym}
	\acro{DER}[DER]{distributed energy resource}
	\acroplural{DER}[DERs]{distributed energy resources}
	\acro{PE}[PE]{power electronic}
	\acro{SMPS}[SMPS]{switch-mode power supply}
	\acroplural{SMPS}[SMPS]{switch-mode power supplies}
	\acro{EMT}[EMT]{electro-magnetic transient}
	\acro{PFC}[PFC]{power factor correction}
	\acro{DUT}[DUT]{device under test}
	\acro{ASD}[ASD]{adjustable speed drive}
	\acro{EMI}[EMI]{electromagnetic interference}
	\acro{THD}[THD]{total harmonic distortion}
\end{acronym}

\usepackage[]{fancyhdr} %
\newcommand{\changefont}{\fontsize{9}{9}\selectfont}
\IEEEoverridecommandlockouts
\begin{document}

\title{Power Response and Modelling Aspects of Power Electronic Loads in Case of Voltage Drops}

\author{\IEEEauthorblockN{Sebastian Liemann, Christian Rehtanz}
\IEEEauthorblockA{Institute of Energy Systems, Energy Efficiency and Energy Economics (ie$^3$),\\
TU Dortmund University, \\Dortmund, Germany \\
*sebastian.liemann@tu-dortmund.de}
}
\maketitle
\thispagestyle{fancy}
\pagestyle{fancy}

\begin{abstract}
In this paper, the power response of power electronic loads in case of voltage drops are measured and their dynamics are analysed. Based on this, dynamic simulation models are derived which can be used for voltage stability investigations. For this, four loads with different power factor techniques are considered. In addition, the impact of the grid impedance and input filter on the power response are measured in the laboratory. Based on the measurements, the simulation models are described. It is also outlined under which aspects the components of the loads are normally dimensioned if no detailed information is available. A comparison with the measurements demonstrates that the simulation models capture the main dynamics. At the end of the paper, the load models are compared to a constant power load in a short-term voltage stability use case. The results indicate that the power electronic loads have a more positive influence on short-term voltage stability in case of voltage drops. Overall, the contributions of the paper are the identification of the basic power dynamics of power electronic loads for different voltage drops and a subsequent derivation of suitable simulation load models for voltage stability investigations.
\end{abstract}


\IEEEpeerreviewmaketitle

\section{Introduction}
In recent years the share of \ac{PE} interfaced loads has steadily increased. This is driven by the advances in semiconductor and \ac{PE} components, resulting in a wide range of devices like light-emitting diodes (LED), adjustable speed drives or \acp{SMPS} for communication technologies and consumer electronics. A survey in the United Kingdom shows that already 30\,\% of the residential load are \ac{PE} loads\,\cite{Tsagarakis2013}. Especially for short-term voltage stability, load dynamics play an important role as they have a decisive influence on the course of the voltage\,\cite{VanCutsem2008}. For this reason, accurate modelling of these upcoming load types is crucial for accurate power system analysis\,\cite{Cigre2014}. However, industry practice shows that a majority uses static load models for dynamic power system studies, like constant power and other conventional load types\,\cite{Milanovic2013}. As shown in\,\cite{Roos2020,Rylander2010,Liemann2021}, \ac{PE} loads do not act like constant loads in case of voltage changes, despite their fast power adjustment. This is especially the case for large-signal stability, for example in the event of short circuits. 
Thus, it becomes clear that the impact of \ac{PE} loads on short-term voltage stability has to be evaluated in more detail, especially if its share will increase in the future. For this reason, accurate dynamic \ac{PE} load models are required to represent their power response for large voltage sags in bulk power system stability studies. In this context, it should be kept in mind that for stability studies a balance must be found between modelling complexity and computational effort. \\
Another aspect is that many dynamic simulations of voltage stability studies assess electromechanical transients\,\cite{Milano2010}. Especially for short- and long-term voltage stability, almost all simulations are done in this phasor or RMS domain\,\cite{Cigre2017}. Yet, it has to be analysed whether the RMS domain is capable of capturing the dynamics of \ac{PE} loads.

By looking into the literature it becomes clear that \ac{PE} loads can be modelled very differently. For power system studies, \ac{PE} loads can be either modelled for steady-state\,\cite{Cresswell2007,Cresswell2008,Cresswell2008a}, small-signal stability\,\cite{Emadi2004}, RMS\,\cite{Ramasubramanian2017,Liemann2021} or \ac{EMT} simulations\,\cite{Roos2020,Collin2020,Collin2011,Collin2010}. In addition, there are also specialised models, e.g. describing the power response in case of voltage sags as a time-domain function\,\cite{Rylander2010}. 
In detail, the authors in\,\cite{Cresswell2007,Cresswell2008,Cresswell2008a} derive generic equivalent circuits from detailed circuits of adjustable speed drives from which they determine steady-state ZIP and exponential load models. Since the focus is on steady-state voltage dependency, no statement can be made according to the power response in case of large voltage drops.
A state-space averaging model of \ac{PE} loads is carried out in\,\cite{Emadi2004} to investigate the small-signal stability in a distribution grid. Here, it is also unclear how these loads affect the large-signal stability of the system. 
A \ac{PE} load model for RMS simulation has been developed in\,\cite{Ramasubramanian2017}. In this case, the authors concentrate on modelling a positive sequence induction motor speed control drive. Several disturbances like load, frequency or voltage changes have been investigated, but the disturbances are rather small. 
In contrast to this, in\,\cite{Collin2020,Collin2011,Collin2010} different types of \ac{PE} loads like LED, electrical vehicle chargers and \acp{SMPS} are measured and also equivalent circuits are developed to match the measurements. In these references, the authors concentrate on the harmonic emission of the \ac{PE} loads in steady-state conditions. Although the match between measurement and simulation is very high in these references, their large-signal behaviour cannot be derived. 
A promising approach is presented in\,\cite{Roos2020} where detailed \ac{PE} load models are compared to laboratory measurements under large disturbances. The authors investigated different light loads, a \ac{SMPS} with active \ac{PFC}, a drive model and two \acp{DER}. As the overall match between measurement and simulation is very good, only one voltage drop of $0.5$\,pu is considered. Here, the authors state that the models have a fewer accuracy if the voltage drop is lower. This is because many \ac{PE} loads disconnect from the grid, if the voltage is too low\,\cite{Yamashita2011}. 

In summary, the literature review reveals that \ac{PE} loads are very diverse, which is also reflected in a large number of different simulation models. However, there is a research gap in \ac{PE} load models which can be used for a wide variety of voltage drops. In order to design such a model, the dynamics of real PE loads must be determined beforehand. Moreover, non-intuitive characteristics can often only be found when testing real loads. This is done by measuring their power response in the laboratory under different voltage drops which cover the desired fault spectrum.

On the one hand, the focus of this paper is to measure and analyse how different \ac{PE} loads respond in case of different voltage sags. On the other hand, suitable load models shall be derived that capture the main dynamics. Since there is a wide variety of different PE loads, this paper concentrates on \ac{SMPS} loads with different \ac{PFC} techniques.  
Moreover, the aim of this paper is to determine the main structure of these loads, but not their parameters. This is because a large number of loads would have to be measured in order to make general statements about their parameters. Nevertheless, with the help of the literature, it is explained how the individual components could be dimensioned if no detailed information or measurements are available. It should also be noted that (short-term) voltage stability is mostly associated with the balance of fundamental reactive and active power. Therefore, this paper also deals mainly with the fundamental component of the power.
Thus, this paper focuses on describing the modelling aspects of \ac{PE} loads in order to capture their fundamental power response in case of voltage drops. 

The paper is structured as follows. In Section\,II real \ac{SMPS} loads are measured in the laboratory to investigate their power response in case of different voltage drops. Here, also the influence of different grid impedances and input filter is analysed.
Afterwards, in Section\,III it is described how these loads can be modelled  to capture their main dynamics. Also, a comprehensive overview is given of how the components can be selected for each model if no detailed information is available. Subsequently, the models are parameterised and a comparison against the measurements is given.  
In Section\,IV, a use case of the simulations load models for short-term voltage stability is given, where they are also compared to a constant power load. The paper closes with a conclusion and outlook in Section\,V. 
\paragraph*{Notation}{In this paper, capital letters like voltage $U$ denote RMS SI values, whereas $U(t)$ denote time-dependent SI values like $U(t) =\hat{U}\cdot sin(\omega\cdot t)$. Small letters $u$ denote per units (pu).}

\section{Measurement of the power response of power electronic loads under voltage drops}
\begin{figure}[b!]
    \centering
    \includegraphics[width=1\columnwidth]{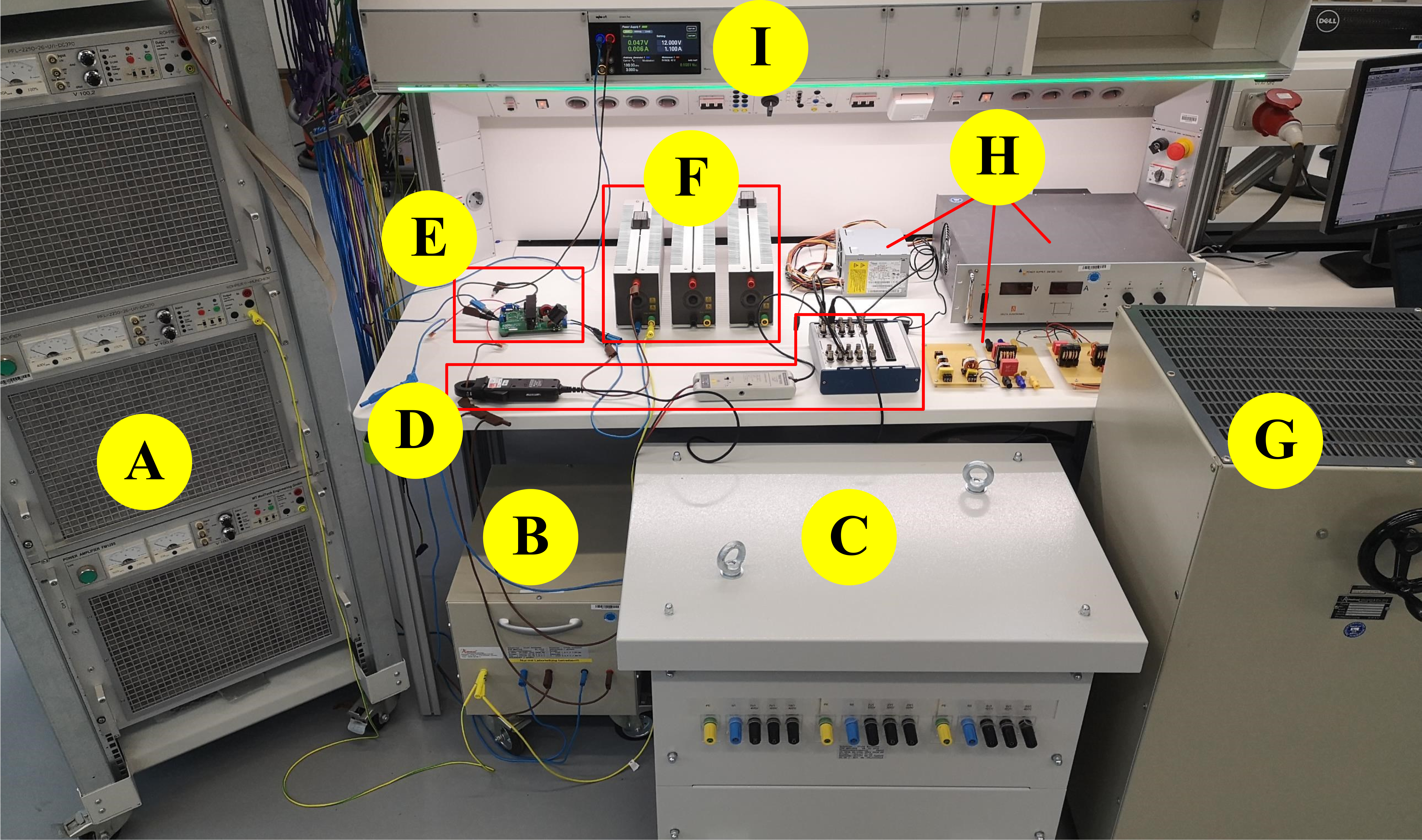}
    \captionsetup{justification=centering}
    \caption{Laboratory setup for measuring the power response of 1ph and 3ph \ac{PE} loads for different voltage drops and periods}
    \label{fig:lab}
    \vspace{0.1cm}
    \includegraphics[width=1\columnwidth]{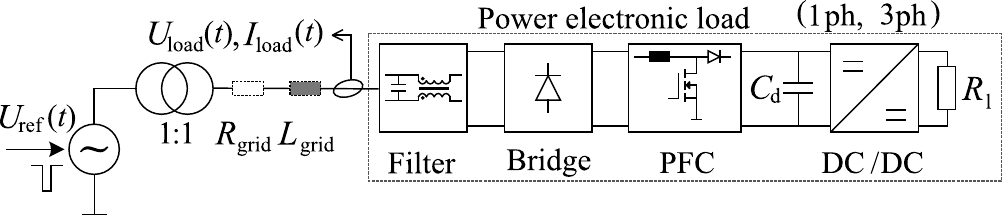}
    \captionsetup{justification=centering}
    \caption{Single line diagram of laboratory setup and illustration of main internal components of the \ac{PE} loads}
    \label{fig:esblab}
\end{figure}
In this section, first, an overview of the laboratory setup is given, which is used to measure the power response of the \ac{PE} loads. Second, the measurement results are presented and the dynamics of each \ac{PFC} type are described. Here, also the influence of a high grid impedance and the \ac{EMI} filter of the loads on the power response is analysed.

\subsection{Laboratory setup} 
\begin{figure}[t]
    \centering
    \includegraphics[width=1\columnwidth]{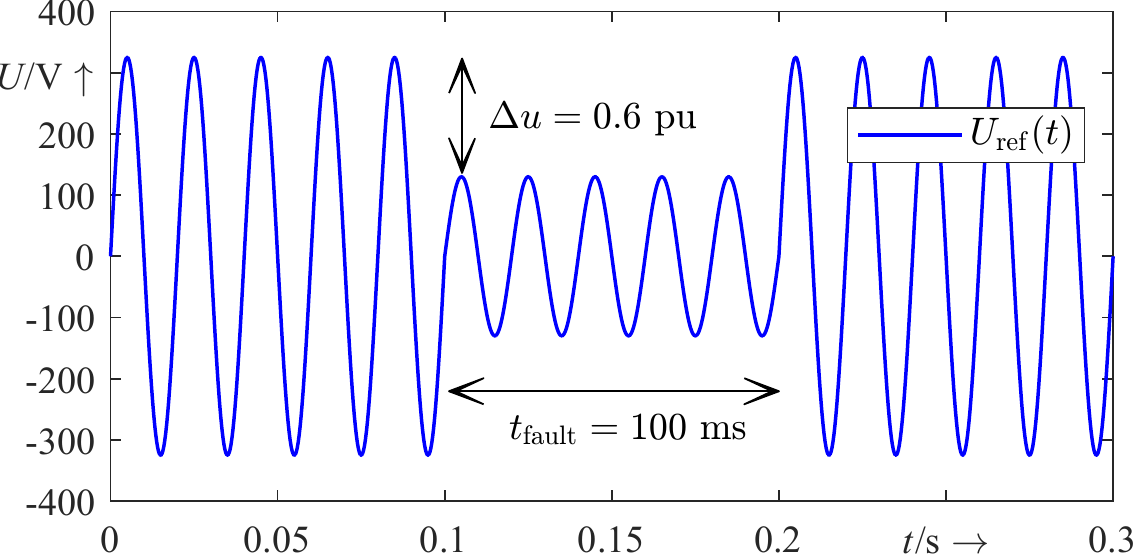}
    \captionsetup{justification=centering}
    \caption{Exemplary voltage reference signal with 60\,\% voltage drop and a fault duration of $100$\,ms}
    \label{fig:vdrop}
\end{figure}

In Fig.\,\ref{fig:lab} the laboratory setup is shown, which is capable of measuring the power of single- and three-phase \ac{PE} loads in high resolution. A single-line diagram of the setup is illustrated in Fig.\,\ref{fig:esblab}, which represents the main components. Here, it has to be noted that the grid impedance, consisting of $R_\mathrm{grid}$ and $L_\mathrm{grid}$, are only used in Section\,\ref{sec:grid_imp}.
For generating the voltage drops, three power amplifiers (Rohrer PFL-2250-26-U/I-DC370) are used, which act like a programmable voltage source, where the desired voltage evolution can be specified  (see (A) in Fig.\ref{fig:esblab}). An exemplary voltage reference signal $U_\mathrm{ref}(t)$ with a voltage drop of $\Delta  u = 0.6\,$pu and its recovery after $100\,$ms is shown in Fig.\,\ref{fig:vdrop}. Here, it has to be noticed that the amplifiers are limited to a maximum current of 30 A (peak). They have an internal current protection that decreases the output voltage in case the current goes beyond the maximum current.  
To add a small impedance between the source and the load, a transformer with a voltage ratio of one is inserted, which can be either single- (B) or three-phase (C). Their electrical parameters can be found in Tab.\,\ref{tab:trafo}. Before the \ac{PE} load, current and voltage sensors are installed, from which the active and reactive power of the fundamental components are calculated (D). The signal recorder has a resolution of $10^5$\,samples per second, which is sufficient for the conducted experiments. Subsequent, the \ac{DUT} is connected, which are the \ac{PE} loads (E). As the \ac{DUT} should operate at rated power, several resistors are available, which can be connected in series or parallel to achieve the desired value (F). For the three-phase loads, there is also a variable resistor with a higher rated current (G). At (H) other \ac{PE} loads and \ac{EMI} filters can be seen. Furthermore, for onboard controllers of a-PFC loads, a $12\,$V DC auxiliary power supply is also provided (I). In this paper, the following \ac{PE} loads are analysed, which are labelled with 'PEL' and a number for convenience:
\begin{itemize}
    \item PEL-1: Coming Data Lithium Battery Charger C2920 
    \item PEL-2: Fujitsu PC Power Supply NPS-230EB B 
    \item PEL-3: PFC Evaluation Board UCC28180EVM-573
    \item PEL-4: DC Power Supply SM300-10D
\end{itemize}
These loads have been selected to have a bright variety of \ac{SMPS} loads with different \ac{PFC} techniques as well as phases. By this, it is possible to determine the different dynamics of each \ac{PFC} circuit on the power response during voltage transients. In\,\cite{Collin2010} and\,\cite{Waniek2017} it has been determined that nearly all \ac{SMPS} can be divided into three \ac{PFC} categories: (1) without or none \ac{PFC} (n-PFC), (2) passive \ac{PFC} (p-PFC) and active \ac{PFC} (a-PFC). In addition, these \ac{PFC} circuits represent the most common in mass-market devices\,\cite{Waniek2017}. Detailed information about each load can be found in Tab.\,\ref{tab:peldata}.

\begin{table}[b]
\caption{Electrical parameters of the transformers}
\begin{tabular}{ccccc}
        & \begin{tabular}[c]{@{}c@{}}1+2 Winding\\ resistance\end{tabular}                                     & \begin{tabular}[c]{@{}c@{}}1+2 Leakage\\ inductance\end{tabular}                   & \begin{tabular}[c]{@{}c@{}}Iron \\ resistance\end{tabular}                                              & \begin{tabular}[c]{@{}c@{}}Primary \\ inductance\end{tabular}                   \\ \hline
B (1ph) & 0.311\,$\Omega$                                                                                      & 0.472\,mH                                                                          & 225.11\,$\Omega$                                                                                        & 1.29\,H                                                                         \\
C (3ph) & \begin{tabular}[c]{@{}c@{}}a: 0.082\,$\Omega$\\ b: 0.079\,$\Omega$\\ c: 0.080\,$\Omega$\end{tabular} & \begin{tabular}[c]{@{}c@{}}a: 0.221\,mH\\ b: 0.226\,mH\\ c: 0.223\,mH\end{tabular} & \begin{tabular}[c]{@{}c@{}}a: 192.39\,$\Omega$\\ b: 532.39\,$\Omega$\\ c: 203.09\,$\Omega$\end{tabular} & \begin{tabular}[c]{@{}c@{}}L1: 1.25\,H\\ L2: 2.45\,H\\ L3: 1.27\,H\end{tabular} \\ \hline
\end{tabular}

\label{tab:trafo}
\end{table}
\begin{table}
\caption{Technical data of the measured \ac{PE} loads}
\begin{tabular}{cccccc}
\multicolumn{1}{l}{} Label & \# Phases & PFC     & $P_\mathrm{r}$ (W) & \begin{tabular}[c]{@{}c@{}}$\mathrm{cos}(\varphi)$\\ meas.\end{tabular} & $U_\mathrm{DC}$ (V) \\ \hline
PEL-1                & 1         & none    & 60        & 0.967 (cap.)                                                & 29.4                \\
PEL-2                & 1         & passive & 230       & 0.937 (ind.)                                                & 12                  \\
PEL-3                & 1         & active  & 360       & 0.996 (cap.)                                                & 390                 \\
PEL-4                & 3         & passive & 3000      & 0.989 (ind.)                                                & 300                 \\ \hline
\end{tabular}

\label{tab:peldata}
\end{table}

\subsection{Power response of PE loads under voltage drops}
\label{sec:power_response}
Next, the measurement results of the \ac{PE} loads are presented. Here, only the fundamental component (50\,Hz) of the active and reactive power are shown, which are determined by calculating the corresponding fundamental current and voltage components by a short-term Fourier transformation. The window length of one Fourier transformation is one period of the fundamental frequency. Due to a sampling frequency of 100\,kHz, this corresponds to 2000 measured values per period. For each new measurement point, a new Fourier transformation is calculated so that the windows of each transformation overlap. 

In Fig.\,\ref{fig:meas_opfc} the power response of the PEL-1 load for different fault duration are shown, where all voltage drops start at $t=0.04$\,s. Furthermore, the voltage drops start at a level of $\Delta u = 0.2\,$pu and increases in steps of $\Delta u = 0.2\,$pu in the next measurement until zero voltage with $\Delta u = 1.0\,$pu.
It has to be noticed, that both active and reactive power are rated to the  corresponding rated active power $P_\mathrm{r}$ of each \ac{PE} load from Tab.\,\ref{tab:peldata}. This makes it easier to compare the power responses of the different loads. In the following, the general power response of a \ac{PE} load during voltage transients is described, using PEL-1 as an example as it is the most simple load. Subsequent, the differences or special characteristics of the other loads with different PFC techniques are highlighted. The subsection concludes with general statements about the dynamics of \ac{PE} under voltage drops.

\subsubsection{General power response of PE loads}
\begin{figure}
    \centering
    \includegraphics[width=1\columnwidth]{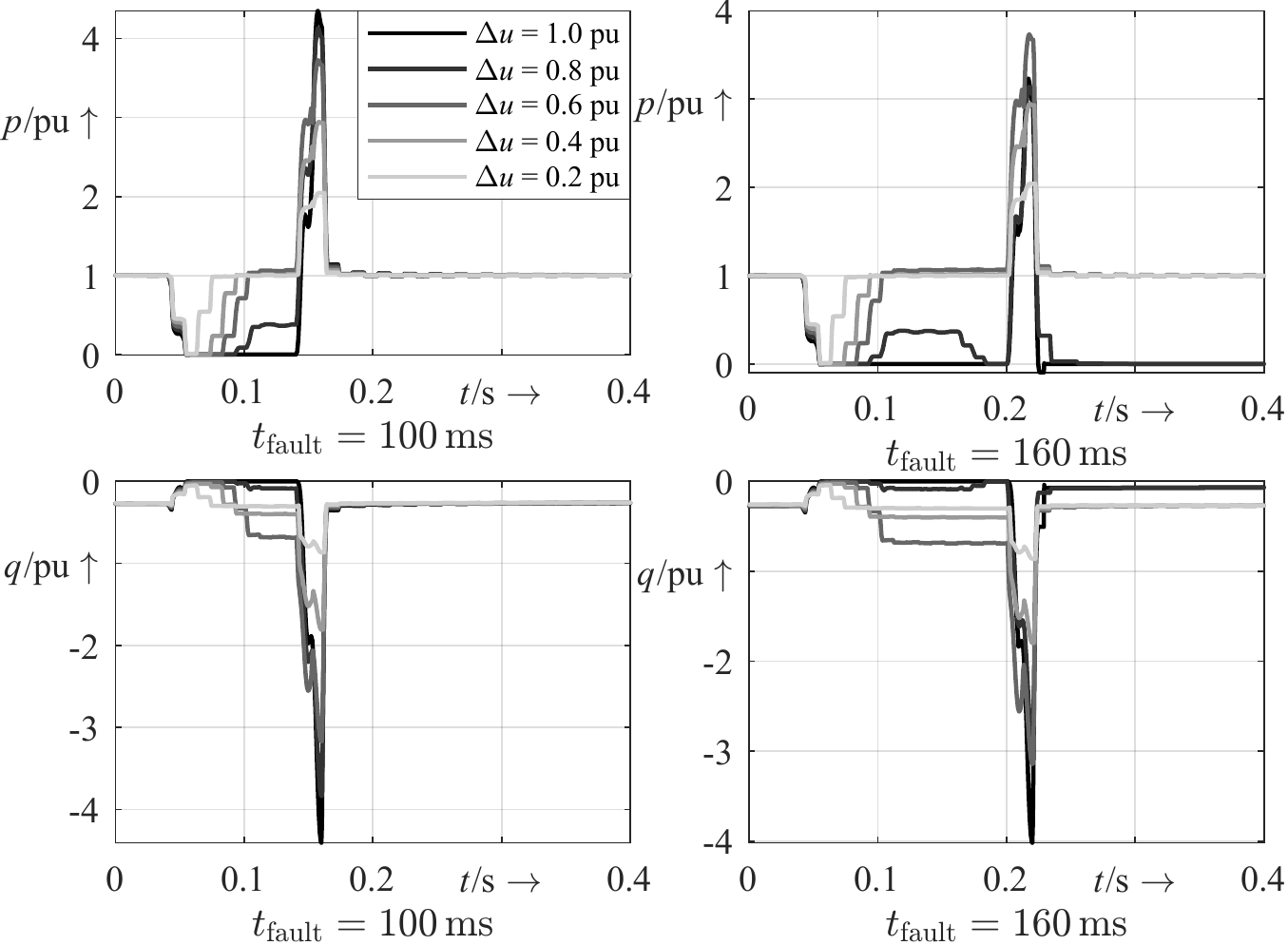}
    \captionsetup{justification=centering}
    \caption{Measured active and reactive power response of PEL-1 for two different fault durations}
    \label{fig:meas_opfc}
\end{figure}

As can be seen in Fig.\,\ref{fig:meas_opfc}, if the voltage starts to drop, the active and reactive power also starts to decline. This is due to the uncontrolled bridge rectifier, disconnecting the load from the grid, as the rectified voltage at the smoothing capacitance $C_\mathrm{d}$ is bigger than the grid voltage. During this time, the actual load $R_\mathrm{l}$ is supplied by the capacitor. If the rectified voltage has decreased low enough, a current starts to flow again into the load, leading to recovering power consumption. The recovery time is longer, if the voltage drop is higher, as the capacitor voltage has to decrease to the value of the grid voltage. Therefore, for zero voltage at $\Delta u=1.0$\,pu no recovery is possible during the fault. For PEL-1, the active power reaches its nominal value during the voltage drop, as an internal DC-DC converter keeps the voltage at the load $R_\mathrm{l}$ constant. Thus, a higher current from the grid flows into the load, compared to normal operation. If the voltage recovers, a high inrush current flows, as the capacitance $C_\mathrm{d}$ is charged during this time. This results in a short but up to four times higher active and reactive  power consumption. It can be concluded that a higher voltage difference at recovery means a larger power consumption afterwards since the capacitor has to be charged more. At the end, the load is in a steady-state again. 
By comparing the power response between the different fault duration, it becomes apparent that fault duration does not lead to a significant change in the evolution. This is because the load recovers during the fault, leading to an intermediate quasi-steady-state. However, as can be seen in Fig.\,\ref{fig:meas_opfc} for $t_\mathrm{fault}=160$\,ms and $\Delta u=0.8$\,pu, the load disconnects during fault and do not recover its power consumption. Moreover, the load disconnects again after recovery and connects itself after $1.4$\,s, which is not shown here. This is also the case for $\Delta u=1.0$\,pu. The reason for this behaviour is not known, but this observation shows that if the voltage drops are deep and long enough, the loads can switch off during and after the fault. Since this effect could be already measured for PEL-2 to PEL-4 either for a fault duration of $t_\mathrm{fault}=100$\,ms or not at all for a longer fault duration, only the power responses for $t_\mathrm{fault}=100$\,ms are shown in the following. Besides, longer fault duration does not show any other characteristics or power response of these loads. In the case of shorter fault duration, the biggest difference is that the power consumption after the fault is smaller. This can be explained by the fact that the internal DC voltage does not drop as low and the capacitor is not charged as much.  
\subsubsection{PFC dependent power response}
\begin{figure}
    \centering
    \includegraphics[width=1\columnwidth]{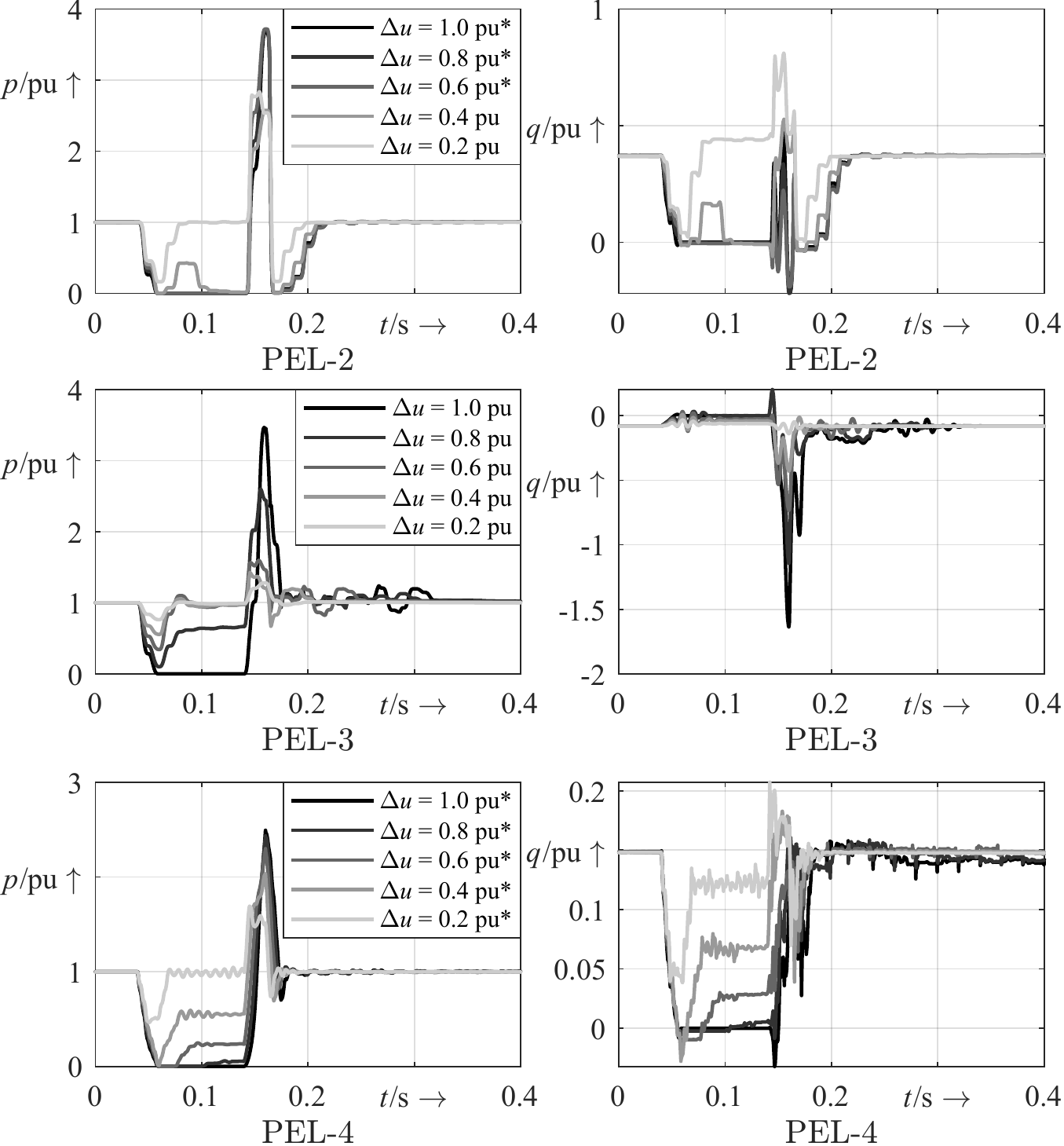}
    \captionsetup{justification=centering}
    \caption{Measured active and reactive power response for $t_\mathrm{fault}=100$\,ms for PEL-2 to PEL-4}
    \label{fig:meas_pel24}
\end{figure}

\begin{figure}
    \centering
    \includegraphics[width=1\columnwidth]{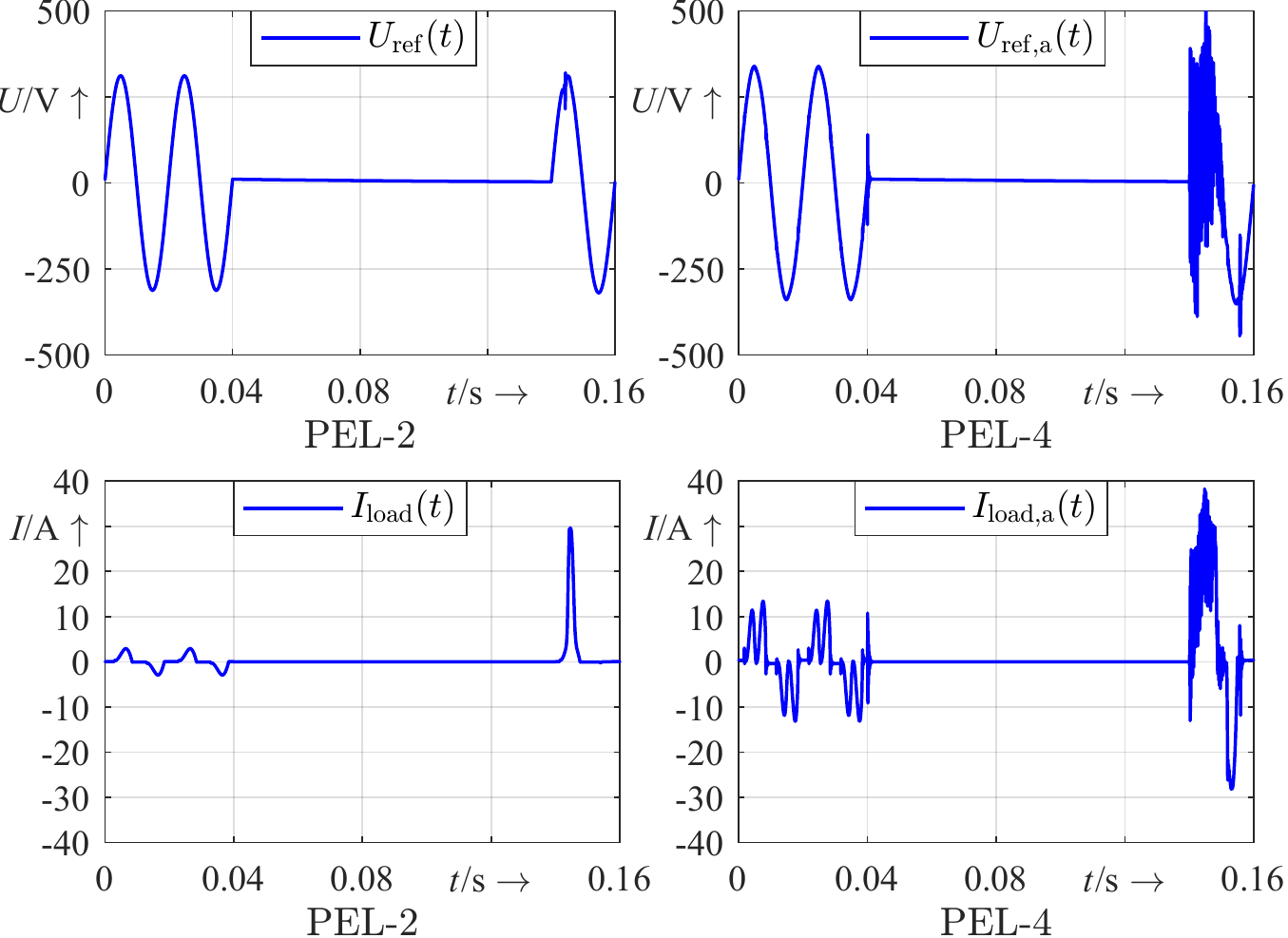}
    \captionsetup{justification=centering}
    \caption{Examples of measured distorted power amplifier voltage due to internal current limitation after $\Delta u =1.0$\,pu}
    \label{fig:screwed}
\end{figure}

Fig.\,\ref{fig:meas_pel24} shows the measurement results of the active and reactive power response for PEL-2 to PEL-4 for a $100$\,ms fault duration. The legend entries apply here per load. The measurements which are marked with an asterisk symbol (*) mean that the power amplifiers had to decrease their voltage, as the requested load current goes beyond their maximum current. However, this only happens after the fault and during voltage recovery due to the high inrush current. Examples of measured distorted amplifier voltages can be seen in Fig.\,\ref{fig:screwed}. Moreover, the degree of voltage distortion varies depending on the load. Therefore, the measurement results of the high power consumption after voltage recovery may not reflect the actual load dynamics. Nevertheless, the measurement results during the fault are not affected. In the case of ideal voltage recovery, higher power consumption could be assumed. Yet, this statement is discussed more in Section\,\ref{sec:tune}.

For PEL-2, the measurement results show that it disconnects already at a voltage drop of $0.4$\,pu. Due to this disconnection, the power evolution for higher voltage drops does not change significantly. In addition, after voltage recovery, there is a short phase where the power consumption is decreased again. This is because a current still flows even after the peak voltage, due to the filtering inductance. As a result, the rectified voltage is higher then the grid peak voltage and has to decline before a load current can flow again.  Compared to PEL-1, the reactive power of PEL-2 is inductive, which can also be explained by the inductance for passive \ac{PFC}. Although the peak current of the PEL-2 is only 3\,A in normal operation, the load tries to draw a slightly larger current than 30 A when the voltage recovers (see Fig.\,\ref{fig:screwed}). This is why the amplifier voltage is shortly decreased at this peak.

In contrast, the current limit is not reached for PEL-3, even though its rated power is higher. The reason is that its active \ac{PFC} reduces the inrush current. Besides, as seen in Fig.\,\ref{fig:meas_pel24}, the internal control leads to a fast power recovery during the fault.  For $\Delta u = 0.8\,$pu it is assumed that the internal boost converter has reached its control limit, leading to a decreased DC voltage and no complete active power restoration. Compared to all other \ac{PE} loads, its reactive power consumption is very low before and during the voltage drop but increases greatly at voltage recovery. Also, slight power  oscillations can be observed afterwards.

In the case of PEL-4, all three power amplifiers reach their current limit during voltage recovery for all voltage drops (see Fig.\ref{fig:screwed} for one phase). This can be explained by the general high power consumption of 3\,kW of this load, which already utilises the amplifiers well. Therefore, the PEL-4 post-fault power response has to be treated with caution and probably does not reflect the true response. It has to be noticed that the active and reactive power of PEL-4 in Fig.\,\ref{fig:meas_pel24} are the sum over all three phases.
However, during the fault, the amplifiers do not reach their limit, but the active power is decreased during normal operation, as shown by Fig.\,\ref{fig:meas_pel24}. During this time, the active power of PEL-4 has a quadratic voltage dependency like a purely resistive load. Here, it is assumed that the internal DC/DC voltage controller has reached its limit and can not draw a higher current, although the amplifiers have not reached their limit yet.
\subsubsection{General dynamics of PE loads during voltage drops}
The measurements demonstrate that all \ac{PE} loads undergo a transient power response in case of significant voltage drops and subsequent recovery. In general, the power response can be divided into three phases. The first phase is the power reduction, as the load is practically disconnected from the grid, due to the blocking rectifier. The second phase is the power restoration if the rectified voltage drops to the grid voltage during the fault. The height of power restoration depends on the limits of the internal DC-DC converter, determining the intermediate steady-state. Also, it can occur that the voltage is too low, leading to a second disconnection. The last phase is during voltage recovery, resulting in a short but high inrush current and corresponding high power consumption, which can be four times as high. For PEL-2 (and partly for PEL-4) a subsequent power reduction is also observed, due to the inductance for passive \ac{PFC}. As the active power dynamics are quite similar for all \ac{PE} loads, the reactive power dynamics vary more. Here, the implemented \ac{PFC} determines whether the reactive power consumption is capacitive (PEL-1 and PEL-3) or inductive (PEL-2 and PEL-4). Based on these insights, the main components and other modelling aspects are derived to develop \ac{PE} load models that can be used in power system stability studies. Before, the influence of the grid impedance and input filter on power response is measured and analysed.

\subsection{Impact of high grid impedance on power response}
\label{sec:grid_imp}
In the following, the impact of a high grid impedance on the \ac{PE} load power response is analysed. The background is that in voltage-critical situations, high power often flows from a generation centre to a load centre, with heavily loaded lines in between. The result is that a high voltage drop occurs over these lines, leading to a reduced voltage at the load side. From the perspective of the load, this circumstance can also be seen as there is a high grid impedance between load and generation. Therefore, it should be analysed how the power response differs compared to normal grid conditions, like in Section\,\ref{sec:power_response}.

To emulate this in the laboratory, a grid impedance $Z_\mathrm{grid}$  consisting of $R_\mathrm{grid}$ and $L_\mathrm{grid}$ ($X_\mathrm{L,grid}$) is inserted between the transformer and the load. The values of the resistance and inductance have been chosen according to two objectives. The first objective is that the approximate voltage drop from the source to the load should be around 10\,\% at steady-state. In this case, the loads operate at their lower voltage band. Since the loads are operating with their rated current, this current shall cause this voltage drop. Therefore, a rated grid impedance $z_\mathrm{grid}$  of 0.1\,pu is needed, based on the rated power $P_\mathrm{r}$ (see Tab.\,\ref{tab:peldata}) and voltage $U_\mathrm{base}=230\,$V of the loads. This can be expressed as follows:
\begin{equation}
    z_\mathrm{grid} = \frac{Z_\mathrm{grid}}{Z_\mathrm{base}} = \frac{Z_\mathrm{grid}}{\frac{U_\mathrm{base}^2}{P_\mathrm{r}}}  = 0.1\,\mathrm{pu}
    \label{eq:zgrid}
\end{equation}
The second objective is that the ratio between inductance and resistance shall reflect the conditions that are usual in low voltage grids. Here, a value around 0.4 is chosen. Therefore, this second condition has to be met:
\begin{equation}
    \frac{X_\mathrm{L,grid}}{R_\mathrm{grid}} = 0.4
    \label{eq:XR}
\end{equation}
These equations can be coupled by calculating the magnitude of the grid impedance $Z_\mathrm{grid}$:
\begin{equation}
    Z_\mathrm{grid} = \sqrt{X_\mathrm{L,grid}^2 + R_\mathrm{grid}^2}
    \label{eq:zg}
\end{equation}
By inserting (\ref{eq:XR}) and (\ref{eq:zg}) in (\ref{eq:zgrid}), the corresponding values for resistance and inductance of each \ac{PE} load can be calculated. However, since only concrete components can be used in the laboratory, deviations occur. Tab.\,\ref{tab:grid_imp} shows the actual values that are used for the measurements. Overall, the real values are close enough to the targeted values. It has to be noted that for PEL-4 no sufficient grid impedances are available at the laboratory, due to the high current requirements. For this reason, the measurements are carried out for PEL-1 to PEL-3 only.

\begin{figure}
    \centering
    \includegraphics[width=1\columnwidth]{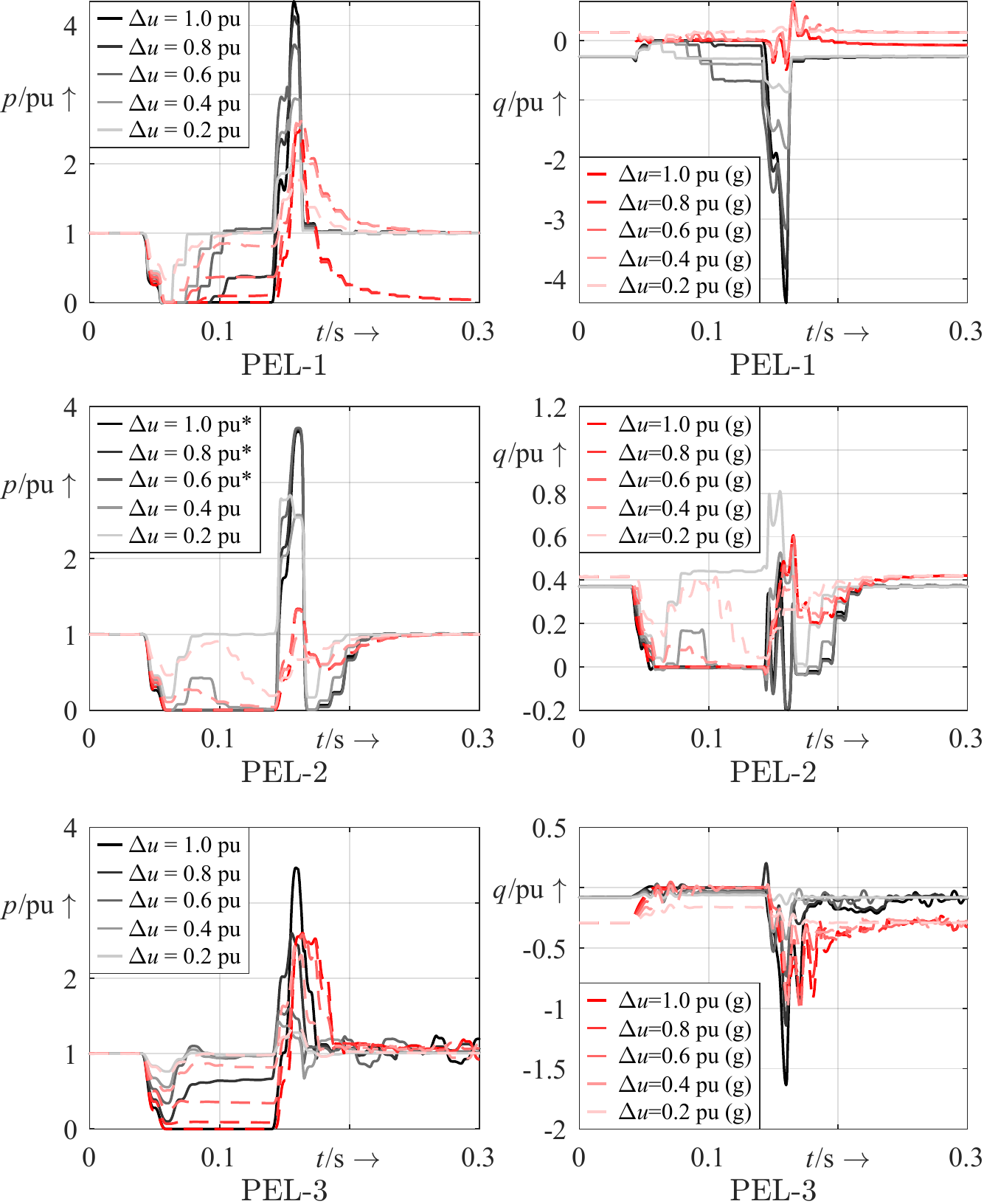}
    \captionsetup{justification=centering}
    \caption{Comparison between the measured active and reactive power response for no and high grid impedance. Measurements with grid impedance are marked with (g).}
    \label{fig:comp_grid_imp}
\end{figure}

Fig.\,\ref{fig:comp_grid_imp} illustrates the comparison of the active and reactive power response between the measurements without (which are the former measurements) and with a grid impedance, marked by a (g). For reasons of space, the legend entries are distributed between the active and reactive power plots, but apply together per load. As shown by the measurements, the grid impedance has a strong impact on all power responses. Moreover, for all measurements with a grid impedance, the current limitation of the power amplifier is never reached as the grid inductance limits the rising current.
A big change can be seen by the greatly reduced power consumption during voltage recovery. One reason is that due to the inrush current, also a big voltage drop occurs over the grid impedance, thus decreasing the voltage at the load. In addition, the grid impedance limits the inrush current.
Looking at PEL-1, the reactive power changes from capacitive to inductive, nearly over the entire duration. This includes also its steady-state reactive power consumption. This can be explained by the fact that the current now has a significantly larger phase shift due to the comparable large grid inductance. However, this is not the case for the reactive power of PEL-3, which is here more capacitive. A reason could be that in this case the grid inductance only serves as bigger storage for the active \ac{PFC} controller.
For active power, however, the pre- and post-fault values are always at the rated power for all loads. In the case of PEL-1 for $\Delta u = 0.8\,$pu and $\Delta u = 1.0\,$pu, the load disconnect itself after voltage recovery, but reconnects after one second, which is not shown here. 
Generally, it can be observed that in case of a high grid impedance the loads are more often unable to recover their active power during the fault or that they switch off earlier. This can be explained by the DC/DC converters and potential protection systems which reach their limits earlier due to the lower voltages.
\begin{table}[]
\caption{Grid impedance values for each \ac{PE} load}
\centering
\begin{tabular}{cccccc}
Label & $R_\mathrm{grid}$ & $L_\mathrm{grid}$ & $X_\mathrm{L,grid}$ & $z_\mathrm{grid}$ & $\frac{X_\mathrm{L,grid}}{R_\mathrm{grid}}$ \\ \hline
PEL-1 & 81.4\,$\Omega$    & 107.73\,mH        & 33.84\,$\Omega$     & 0.100\,pu           & 0.416                                     \\
PEL-2 & 22.8\,$\Omega$    & 29.71\,mH         & 9.33\,$\Omega$      & 0.107\,pu         & 0.409                                     \\
PEL-3 & 14.4\,$\Omega$    & 20.81\,mH         & 6.54\,$\Omega$      & 0.108\,pu         & 0.454                                     \\ \hline
\end{tabular}

\label{tab:grid_imp}
\end{table}

From these measurements, it can be concluded that the high power consumption at voltage recovery is significantly reduced and that power recovery during the fault is not as frequent as before. Also, due to the low voltage, there are earlier disconnections of the load.
These measurements also demonstrate that especially reactive power depends on the grid impedance, even for steady-state.
In addition, the grid impedance can not only be characterised by its causing voltage drop, but also by its inductance dynamics and resulting current phase shift. If the characteristics of the \ac{PE} loads would only be dominated by the voltage magnitudes, similar characteristics should have been noticed from the former measurements. As pointed out in the introduction, most voltage stability investigations are done in the phasor domain. As the observations made here show dynamics faster than the fundamental frequency and the interaction between grid impedance and \ac{PE} load can only be adequately described by the differential equations of inductances and capacitances, it is concluded that the EMT domain should be used for the simulation of these loads.
In summary, this means that a high grid impedance has a strong impact, both on active and reactive power response and that the dynamics of the loads can only be captured in simulations with great detail.

\subsection{Impact of EMI filter on power response}
\begin{figure}[t]
    \centering
    \includegraphics[width=0.8\columnwidth]{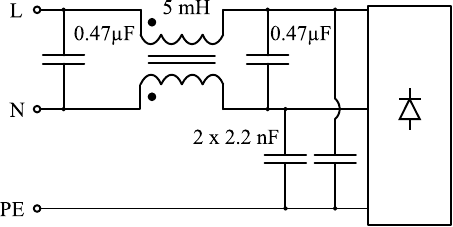}
    \captionsetup{justification=centering}
    \caption{Grid side EMI filter of PEL-3 (own representation, based on\,\cite{Wohlfahrt2017})}
    \label{fig:emifilter}
\end{figure}

\begin{figure}
    \centering
    \includegraphics[width=\columnwidth]{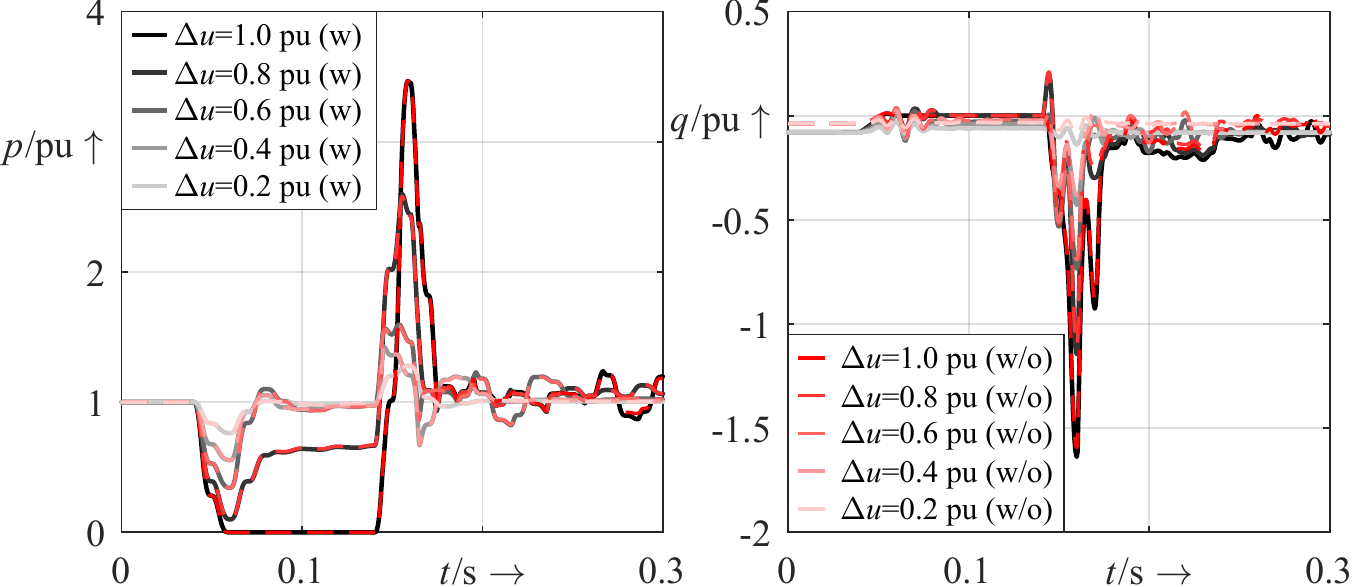}
    \captionsetup{justification=centering}
    \caption{Comparison of measured active and reactive power response of PEL-3 with (w) and without (w/o) EMI filter}
    \label{fig:measemi}
\end{figure}

Next, the impact of the grid side \ac{EMI} filter of the \ac{PE} load on the power response is analysed. For this, an additional measurement is carried out where the \ac{EMI} filter of PEL-3 is unsoldered. Generally, the \ac{EMI} filter shall attenuate high-frequency noise which is emitted by the load, but also to protect the load from high grid harmonics\,\cite{Wohlfahrt2017}. The filter circuit is illustrated in Fig.\,\ref{fig:emifilter}, where the inductance is a common mode choke.  
Fig.\,\ref{fig:measemi} shows the power response of PEL-3 with and without the \ac{EMI} filter. As the results indicate, the filter has nearly no influence on the fundamental active component. For reactive power, only small differences can be seen. For example, the steady-state reactive power consumption decreases from 26.5\,VAr (cap.) to 12.5\,VAr (cap.), which results in a higher power factor from 0.997 to 0.999. However, as can be seen by the numbers, the impact of the filter is almost negligible. In addition, the reactive power evolution during and after the voltage drop is slightly different, but the influence is also small. For this reason, it can be concluded that the \ac{EMI} filter has a negligible impact on the fundamental power response and can therefore be neglected for modelling, although this leads to small errors within the reactive power.


\section{Modelling aspects of power electronic loads in case of voltage drops}
\begin{figure*}[t]
    \centering
    \includegraphics[width=\textwidth]{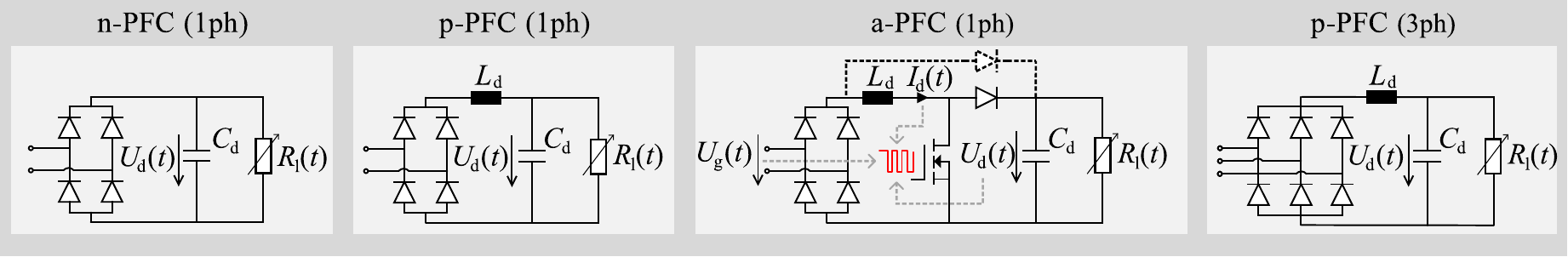}
    \captionsetup{justification=centering}
    \caption{Overview of circuit design for PEL-1 to PEL-4 with different \ac{PFC} techniques}
    \label{fig:pfcoverview}
\end{figure*}
In this section, it is described how the measured \ac{PE} loads have to be modelled to capture their main dynamics for voltage drops, e.g. in short-term voltage stability studies. At first, a general overview of the structure of different \ac{PFC} circuits is given. Here it has to be noted the structures of the loads are greatly inspired by several references, like\,\cite{CresswellDiss,Rylander2010,Cigre2014,Collin2010}.
In this context, also potential default parameters or how to specify them are discussed. Yet, these default parameters should not be considered as actual values. Instead, they are approximations to show how these parameters can be estimated if detailed information on specific loads is not available. Moreover, a simple procedure is presented to reproduce the disconnection of the loads in case of low voltages. Subsequent, the parameters of the load models are tuned and their dynamic response is compared to the measurements. 

\subsection{General structure and component sizing of power electronic loads}
Fig.\,\ref{fig:pfcoverview} illustrates the different circuits of \ac{PFC} techniques which also represent the loads PEL-1 to PEL-4. In contrast to Fig.\,\ref{fig:esblab}, all \ac{PE} models have a variable resistor that acts as a constant power load instead of the DC-DC converter. This assumption has also been made in other studies\,\cite{Cigre2014,CresswellDiss,Rylander2010}, as the DC-DC converter keeps the DC voltage at the load constant, even at larger disturbances. Therefore, the equivalent load resistance $R_\mathrm{l}(t)$ depends on the rectified voltage $U_\mathrm{d}(t)$ and is calculated as follows:
\begin{equation}
    R_\mathrm{l}(t) =
        \begin{cases}
           \frac{U_\mathrm{d}^{2}(t)}{P_\mathrm{r}},& U_\mathrm{d}(t) \geq U_\mathrm{off} \\
           \infty,& else 
        \end{cases}
\end{equation}
As can be seen by the equation, the resistance $R_\mathrm{l}(t)$ also depends on whether the capacitance voltage $U_\mathrm{d}(t)$ is greater than a threshold, which is here the switch-off voltage parameter $U_\mathrm{off}$. So in case the voltage $U_\mathrm{d}(t)$ drops below this value the resistor becomes infinitely large and thus no power is drawn from the grid, leading to a quasi disconnection.
Also, for all \ac{PE} loads an ideal uncontrolled bridge rectifier is assumed, either single- or three-phase.

\begin{figure}
    \centering
    \includegraphics[width=1\columnwidth]{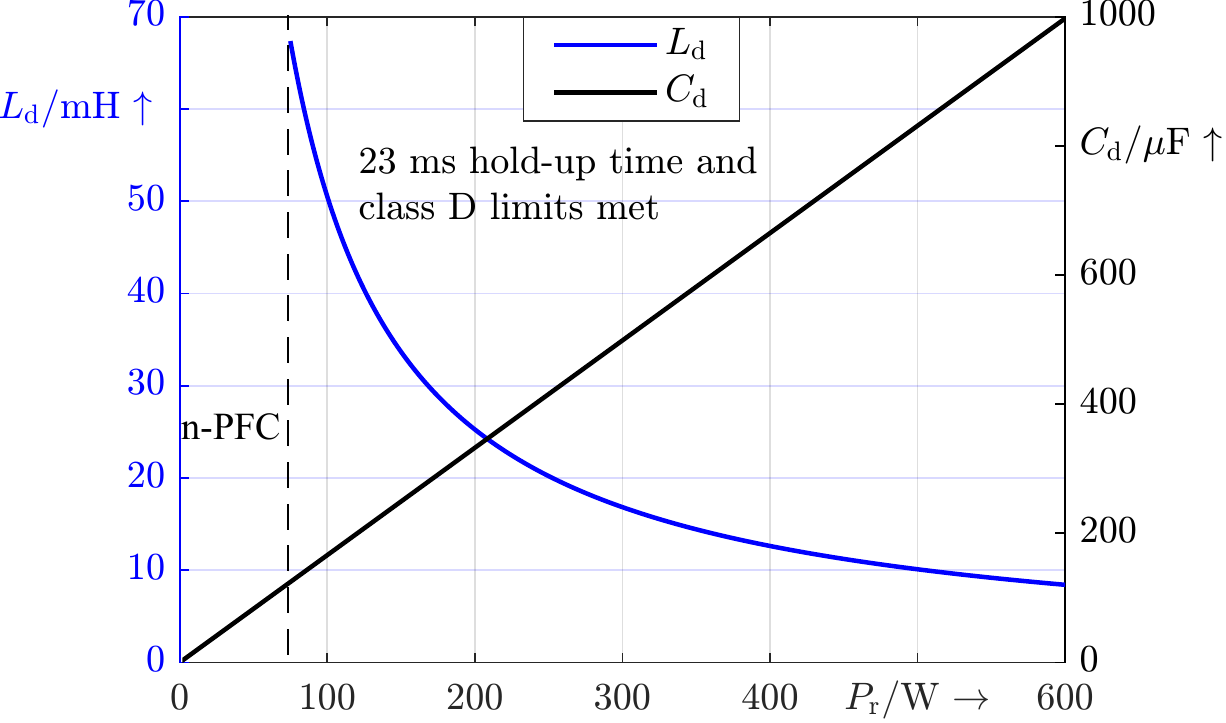}
    \captionsetup{justification=centering}
    \caption{Sizing of smoothing capacitance $C_\mathrm{d}$ to meet 23\,ms hold-up time and filter inductance $L_\mathrm{d}$ to meet class D limits  (own representation, based on \cite{CresswellDiss})}
    \label{fig:sizing}
\end{figure}

For most \ac{SMPS} loads, the design of the smoothing capacity $C_\mathrm{d}$ depends on the rated power, the voltage input range and the hold-up time. The following equation can be used to calculate the needed capacitance\,\cite{CresswellDiss}:
\begin{equation}
    C_\mathrm{d} = \frac{2\cdot P_\mathrm{r}\cdot t_\mathrm{hold}}{\eta\cdot(U_\mathrm{nom}^2 - U_\mathrm{min}^2)}
    \label{eq:cd}
\end{equation}
Here, the time $t_\mathrm{hold}$ is the hold-up time and describes how long the capacitance can supply the load, in case of an undesired disconnection from the grid. A typical value is $23$\,ms which is also the main criteria in sizing this capacitance\,\cite{CresswellDiss}. The parameter $\eta$ is the efficiency of the load and $U_\mathrm{nom}$ is nominal phase to ground RMS voltage and $U_\mathrm{min}$ the minimum input voltage at which normal operation shall be possible for the load. Fig.\,\ref{fig:sizing} shows typical values of $C_\mathrm{d}$ depending on the rated power $P_\mathrm{r}$ with $t_\mathrm{hold} = 23\,$ms, $ \eta = 95\,\%$, $U_\mathrm{nom} = 230\,$V and $U_\mathrm{min} = 150\,$V (values taken from\,\cite{CresswellDiss}). If these values are related by $U_\mathrm{nom}$ and $P_\mathrm{r}$, a per unit impedance $x_\mathrm{Cd,pu}$ for $C_\mathrm{d}$ can be calculated:
\begin{equation}
    x_\mathrm{Cd,pu} = \frac{1}{C_\mathrm{d} \cdot 2\pi\cdot50\,\mathrm{Hz}} \cdot \frac{P_\mathrm{r}}{U_\mathrm{nom}^2} \approx 0.036\,pu
\end{equation}
With this value, the capacitance $C_\mathrm{d}$ can be calculated for a different base power and base voltage, e.g. for a specific load. 

Similarly, a per unit value can be calculated for the filtering induction $L_\mathrm{d}$, which is here for passive \ac{PFC}. The aim of the inductor $L_\mathrm{d}$ is to reduce the load current harmonics. Therefore, its sizing aims to meet
the harmonic limits for class D devices of the IEC norm 61000-3-2\,\cite{IEC61000-3-2}. Here, class D devices range from 75\,W to 600\,W. As the harmonic limits depend on the rated power of the device, a per unit value for $L_\mathrm{d}$ is also possible. In\,\cite{CresswellDiss} the following per unit impedance $x_\mathrm{Ld,pu}$ has been determined for which the harmonic class D limits are met:
\begin{equation}
    x_\mathrm{Ld,pu} =  2\pi \cdot50\,\mathrm{Hz} \cdot L_\mathrm{d} \cdot \frac{P_\mathrm{r}}{U_\mathrm{nom}^2} \approx 0.03\,pu   
\end{equation}
Values of $L_\mathrm{d}$ meeting this requirements are also presented in Fig.\,\ref{fig:sizing}
as a function of the rated power. As for loads smaller than 75\,W no harmonic limits are specified, there is also no need for a filtering inductance and thus no \ac{PFC}.

For active \ac{PFC}, the sizing of $L_\mathrm{d}$ is a bit different, since its primary function is now to work as temporary storage of the current $I_\mathrm{d}$ of the boost converter (see Fig.\,\ref{fig:pfcoverview}). In the following, the dimensioning process is made using the example of the design of load PEL-3\,\cite{UCC} and is, therefore, consistent for the later modelling of the circuit. It should be noted here that other manufacturers may design the active \ac{PFC} slightly differently, depending on the underlying design goals, e.g. continuous or discontinuous current mode\,\cite{Infineon,ONS}.
Here, the inductance $L_\mathrm{d}$ can be calculated by the following equation\,\cite{UCC} :
\begin{equation}
   L_\mathrm{d} \geq \frac{U_\mathrm{nom}\cdot \mathrm{D}(1-\mathrm{D})}{I_\mathrm{ripple}\cdot f_\mathrm{s}}
\end{equation}
In the above equation, D is the duty cycle (which is assumed here as 0.5 for the worst case), $I_\mathrm{ripple}$ the ripple current and $f_\mathrm{s}$ the switching frequency. The inequality of the equation means that $L_\mathrm{d}$ should be at least high as the right-hand side of the equation. The current ripple can be estimated by the following formula, where it is assumed that it is around 40\,\% of the maximum RMS input current:
\begin{equation}
    I_\mathrm{ripple} = 0.4 \cdot \frac{\sqrt{2} \cdot P_\mathrm{r}}{\eta \cdot U_\mathrm{g,min} \cdot \mathrm{PF}}
\end{equation}

\begin{figure}[b]
    \centering
    \includegraphics[width=1\columnwidth]{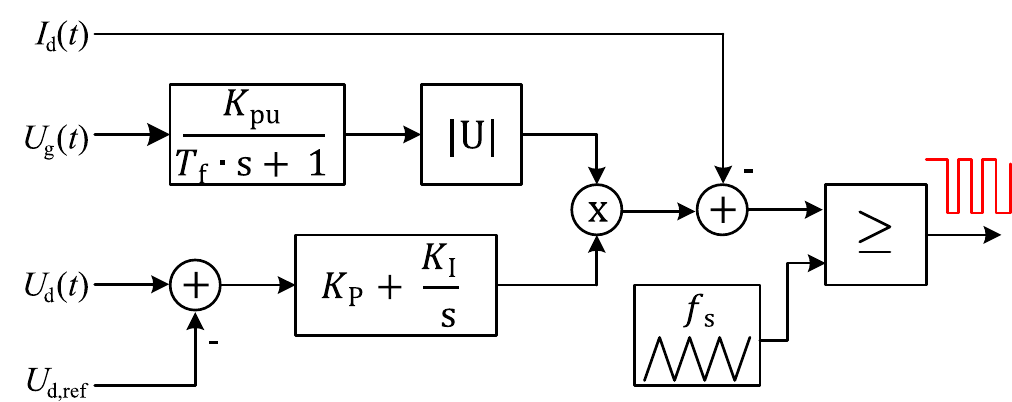}
    \captionsetup{justification=centering}
    \caption{Controller of boost converter for active \ac{PFC}}
    \label{fig:apfccontrol}
\end{figure}

To calculate the current ripple, values for the efficiency $\eta$, the minimum RMS input voltage $U_\mathrm{g,min}$ and the power factor PF have to be specified. As these values are design parameters for the active \ac{PFC}, no further information can be given here. However, in\,\cite{UCC} actual values for these parameters for PEL-3 are provided. The switching frequency $f_\mathrm{s}$ is also assumed there. For active \ac{PFC}, the calculation of $C_\mathrm{d}$ is the same as in (\ref{eq:cd}). To model the active PFC stage or boost converter, an additional switch and one or two diodes are necessary  (see Fig.\,\ref{fig:pfcoverview}). The dashed plotted diode is the so-called pre-charge diode which charges the capacitance $C_\mathrm{d}$ directly after it has been connected to the grid. Generally, this diode is optional but is used for modelling PEL-3 later. To realise the active \ac{PFC}, the ideal switch has to be controlled regarding two control targets\,\cite{Tulay2017} (see Fig.\,\ref{fig:apfccontrol}). First, the voltage $U_\mathrm{d}(t)$  is regulated to the reference value $U_\mathrm{d,ref}$ over a PI-controller. This reference is higher than the peak grid voltage, as a boost converter stage is used. Second, the current $I_\mathrm{d}(t)$ is controlled in a way that it should take on the temporal course of the grid voltage. As a result, the load current and grid voltage are in phase and thus the power factor decrease to a small value. In some applications, an additional current control can also be used\,\cite{Tulay2017}, but is omitted here, due to simplicity. Besides, an additional first-order lag is used to filter high harmonics of the measured grid voltage. Note that the time constant should not be too large in order to avoid a phase shift to the grid voltage. The gain $K_\mathrm{pu}$ is set according to the reciprocal peak value of the grid voltage $\hat{U}_\mathrm{g}$.  

For three-phase passive \ac{PFC} loads, the sizing of the capacitance $C_\mathrm{d}$ and inductance $L_\mathrm{d}$ is different, compared to single-phase passive \ac{PFC} devices. On the one hand, the rectified voltage $U_\mathrm{d}(t)$ is larger by a factor of $\sqrt{3}$ and the capacity $C_\mathrm{d}$ is charged more often per period. On the other hand, the harmonic limits for three-phase loads can vary greatly, depending on their application. In IEC 61000-3-2 symmetrical three-phase loads with a rated current per phase smaller than 16\,A are class A devices, if they do not belong to another class\,\cite{IEC61000-3-2}. For phase currents greater 16\,A and lower 75\,A, IEC norm 61000-3-12 has to be applied\,\cite{IEC61000-3-12}. If the load is an \ac{ASD} and does not belong to IEC 61000-3-2 or IEC 61000-3-12, the harmonic limits are specified by IEC norm 61800-3\,\cite{IEC61800-3}. However, if the three-phase load is used as 'professional equipment' with a rated power greater than 1\,kW no specific harmonic limits are specified\,\cite{IEC61000-3-2}. Here, PEL-4 falls into this last category, as this DC power supply is mostly used in a laboratory environment. This assumption is supported by the fact that measurements from PEL-4 show that under normal conditions the current contain harmonics from the 5\textsuperscript{th} and 7\textsuperscript{th} order which are 2.5 times higher than the respective harmonic limits (cf. Fig.\,\ref{fig:screwed} of PEL-4 before $t=0.04$). Therefore, the PEL-4 inductance has not to be selected to meet these limits and the device can be classified as professional equipment. It can be concluded that for three-phase loads the sizing of $L_\mathrm{d}$ and $C_\mathrm{d}$ is individual and must be considered per device. For that reason, no further specifications can be made here. For those who are interested to know how the sizing is done for \ac{ASD} loads, the reader is referred to\,\cite{Cresswell2008} and \cite{Cresswell2008a}. 

\subsection{Adoption of simulation models and comparison to measurements}
\label{sec:tune}
In this section, a comparison between the measurements and simulation models is made. For this, the model parameters in Fig.\,\ref{fig:pfcoverview} are adopted manually so that they have a similar power response compared to the measurements. However, as mentioned before, the goal is not a perfect replica of the power response, but one that encompasses the main dynamics. For a replication that is as accurate as possible, reference can be made to parameter tuning methods in\,\cite{Roos2020,Tulay2017,Rojas2021}. Generally, for common or generic parameters of \ac{PE} loads, a larger number would have to be measured and the values and distribution of these parameters have to be analysed. Here, the focus is on describing the main dynamics of \ac{SMPS} loads with different \ac{PFC} types and how they can be modelled from a structural perspective.
\subsubsection{Utilised model parameters}
An overview of the utilised values for the simulation models is given in Tab.\,\ref{tab:simdata}. Here, the parameter $u_\mathrm{off}$ for disconnecting the load is based on the peak value of $U_\mathrm{base}$ which is $230\cdot\sqrt{2}\,$V.  It has to be noticed that the values were selected according to the fundamental power response and not to the deformed power or the \ac{THD}. The asterisk symbol\,* in Tab.\,\ref{tab:simdata} means that instead of voltage limit, a minimum value for $R_\mathrm{l}$ has been set. This is because, PEL-4 shows a clear quadratic voltage dependence below a certain voltage value (cf. Fig.\,\ref{fig:meas_pel24} (e)), which results from a constant resistance. The minimum value is set so $63.11\,\Omega$, whereby this value is reached approximately at a voltage of 65\,\% of the rectified voltage $U_\mathrm{d}(t)$. The control parameters for PEL-3 are listed in Tab.\,\ref{tab:apfccontrol} according to the structure from Fig.\,\ref{fig:apfccontrol}, which are also selected manually. In general, it is more likely that the control parameters are selected according the desired power factor, \ac{THD} or response time in case of load steps\,\cite{Tulay2017}. The values of $C_\mathrm{d}$, $L_\mathrm{d}$, $U_\mathrm{d,ref}$ and $f_\mathrm{s}$ of PEL-3 are taken from its data sheet in\,\cite{UCC}. 
\begin{table}[h]
\caption{Electrical values and base parameters of the \ac{PE} simulation models}

\centering
\begin{tabular}{cccccc}
Label & $x_\mathrm{Cd,pu}$ & $x_\mathrm{Ld,pu}$    & $u_\mathrm{off}$ & \multicolumn{1}{l}{$U_\mathrm{base}$} & $P_\mathrm{base}$ \\ \hline
PEL-1 & 0.058 pu           & -                     & 0.0 pu             & 230 V                                          & 60 W                       \\
PEL-2 & 0.06 pu            & 0.034 pu              & 0.65 pu          & 230 V                                          & 230 W                      \\
PEL-3 & 0.1 pu             & $0.0007\,$ pu & 0.1 pu           & 230 V                                          & 360 W                      \\
PEL-4 & 0.361 pu           & 0.0125 pu             & 0.0 pu *           & 230  V                                          & 3000 W                     \\ \hline
\end{tabular} 
\captionsetup{justification=centering}
\label{tab:simdata}
\end{table}
\begin{table}[h]
\caption{Parameters of active \ac{PFC} controller for PEL-3}
\centering
\begin{tabular}{cccccc}

$K_\mathrm{pu}$         & $T_\mathrm{f}$ & $K_\mathrm{P}$ & $K_\mathrm{I}$ & $f_\mathrm{s}$ & $U_\mathrm{d,ref}$ \\ \hline
$\frac{1}{230\sqrt{2}}$ & 0.0005\,s         & 0.08\,pu           & 0.15\,pu           & 165.0 kHz      & 390 V              \\ \hline
\end{tabular}

\label{tab:apfccontrol}
\end{table}
\begin{figure*}
    \centering
    \includegraphics[width= 18cm]{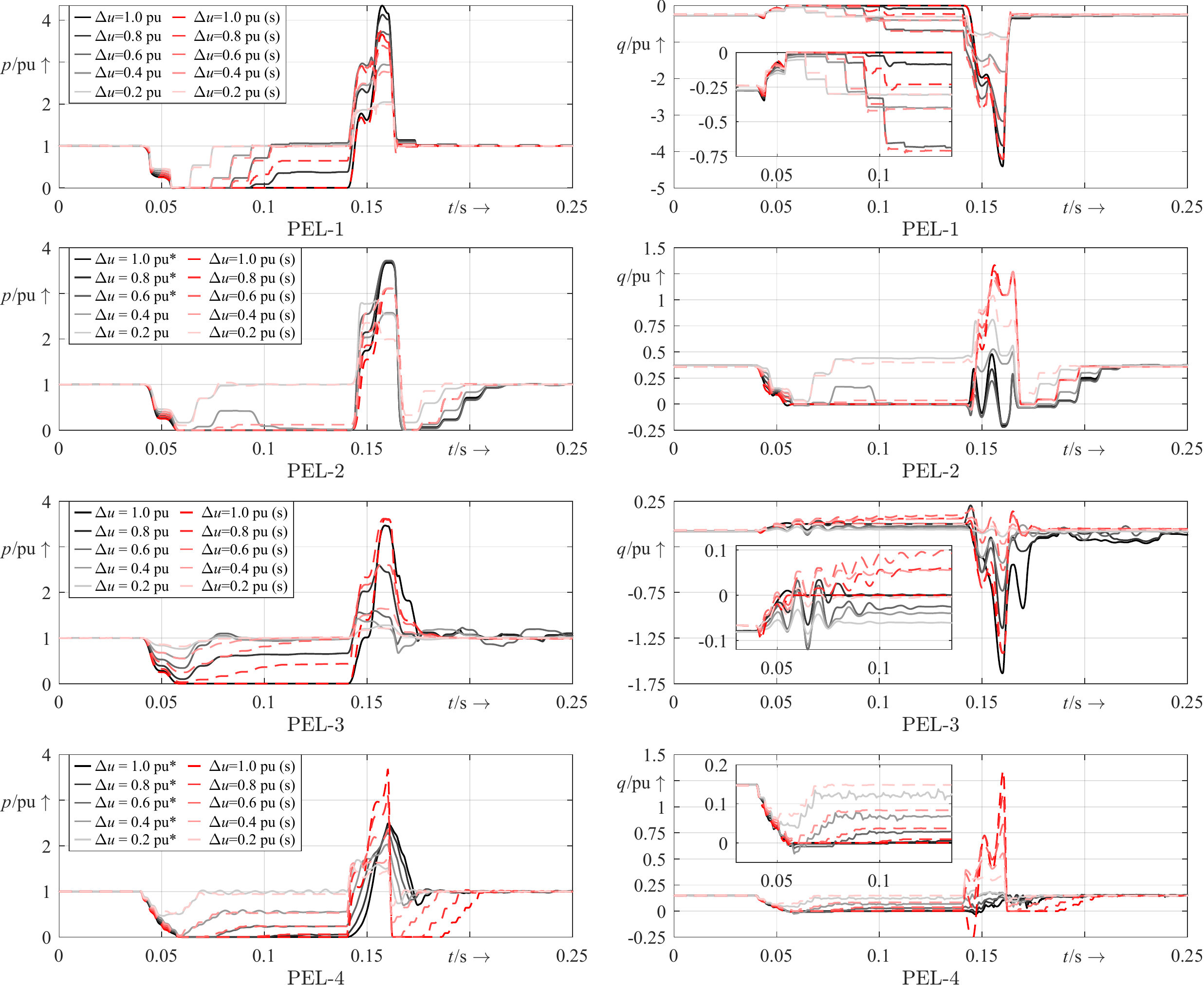}
    \captionsetup{justification=centering}
    \caption{Comparison of active and reactive power response between measurement and simulation for all \ac{PE} loads. Simulation results are marked with an (s). Measurement results marked with an (*) contain current limitation during voltage recovery.}
    \label{fig:compmeassim}
\end{figure*}
\subsubsection{Measurement and simulation comparison}
Fig.\,\ref{fig:compmeassim} shows the comparison of the power response between the measurement and simulation for all \ac{PE} loads. For the simulations, a simple transformer model is used by inserting only the winding resistance and leakage inductance of the laboratory transformer (cf.\,Tab.\,\ref{tab:trafo}). The simulations are carried out in MATLAB Simulink.

As can be seen by the results, the correlation between measurement and simulation for PEL-1 is very high. Here, only small differences are noticeable, like the maximum power consumption during voltage recovery or the active power consumption for $\Delta u = 0.8\,$pu during the fault. For PEL-2, the discrepancies are mainly during voltage recovery. Here, there are small for active but bigger for reactive power. A possible explanation for this could be the current limitation of the amplifiers during voltage recovery. Yet this can only be assumed, although it could explain why the greatest differences occur at the time of current limitation. This conclusion is supported by the simulation results of PEL-3, as there is no current limitation and the correlation is high for active and reactive power.
Here, a slight mismatch regarding the time of power restoration during the fault can be observed. Also, in the simulations, there are no small oscillations around the nominal values after the fault. These mismatches could be explained by the fact that the controller of the active \ac{PFC}  from Fig.\,\ref{fig:apfccontrol} is not the same as used in\,\cite{UCC}. In the last comparison of PEL-4, it can be seen that there is a great mismatch during voltage recovery and shortly after it. Since the distortion of the voltage from the power amplifier is much bigger compared to PEL-2, it is plausible that this circumstance is also reflected here (cf. Fig.\,\ref{fig:screwed}). Before voltage recovery, the correlation between active and reactive power is very high. Fr that reason, it is concluded that the simulation results better describe how PEL-4 would respond in case of a stiffer voltage source. 

In summary, the comparison demonstrates that the simulation models capture the main dynamics of the measured loads. However, due to limitations of the laboratory equipment, there are greater mismatches for PEL-2 and PEL-4. A further source of error could be the transformer model and also the accuracy of the transformer data from Tab.\,\ref{tab:trafo}. In addition, laboratory cables and plugs contain additional impedances that have not been taken into account here.
Since the component and parameter values are selected manually, there could also be  potential for better correlation. However, this is not the objective here. Even if more appropriate parameters would have been found, these values would not represent general values, as they belong to specific loads.

\section{Use case: impact of pe loads on short-term voltage stability}
In the final part of this paper, a simulation use case of the derived load models is shown, in which their influence on short-term voltage stability is investigated. In a previous analysis, it could be demonstrated that \ac{PE} loads have the biggest difference in short-term dynamics, compared to constant power loads\,\cite{Liemann2021}.
For this, a simulation comparison between the four \ac{PE} loads for three different fault scenarios is shown. Moreover, the results are compared against an RMS simulation with a constant power load. Therefore, it can be assessed to what extent the assumption is correct that \ac{PE} loads can be modelled as constant power loads. 

\begin{figure}
    \centering
    \includegraphics[width=1.0\columnwidth]{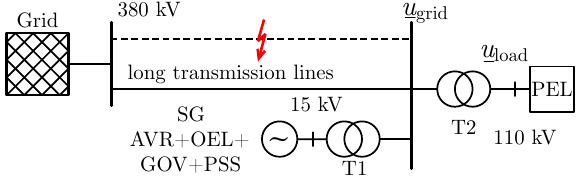}
    \captionsetup{justification=centering}
    \caption{Voltage stability test system (own representation, based on\,\cite{Ospina2020})}
    \label{fig:LTVS_test_system_Svg}
\end{figure}

To do so, a four-bus test system (see Fig.\,\ref{fig:LTVS_test_system_Svg}) is used, where its structure is taken from\,\cite{Ospina2020} and the values of the transmission lines from\,\cite{Nordic}. The test system is a simple representation of a generation and a load area that are connected by two long transmission lines. The simplicity of the test system has been chosen to make it easier to identify the impact of the \ac{PE} loads. Thus, the following investigations can only give an impression of how \ac{PE} loads could respond in a voltage-critical situation.  Detailed parameters about the grid model can be found in Tab. \ref{tab:3bus}. Inside the load area, a synchronous generator is connected which is equipped with a governor (GOV), an automatic voltage regulator (AVR), an over-excitation limiter (OEL) and a power system stabiliser (PSS). In all scenarios, the generator injects its nominal active power and regulates its terminal voltage to 1.0\,pu.
The parameters of the generator and its controller are taken from the generator G2 in\,\cite{Nordic}. To incorporate the single-phase loads PEL-1 to PEL-3, three loads of each type are used, which are arranged in a delta connection. In steady-state, the summarised active power $P_\mathrm{l}$ of these three loads is either $ 500\,$MW or $800\,$MW, depending on the scenario. For PEL-4, only one load is used. Since the reactive power depends on the respective load and grid impedance, no general value can be given here. The simulations are carried out in MATLAB Simulink.

At the beginning of each simulation, a three-phase fault with a fault resistance $R_\mathrm{f}$ occurs in the middle of the upper transmission line, which is cleared by disconnecting this line. For all simulation results figures, active and reactive power are related to the corresponding rated active power $P_\mathrm{l}$ of the load.

\begin{figure}
    \centering
    \includegraphics[width=1\columnwidth]{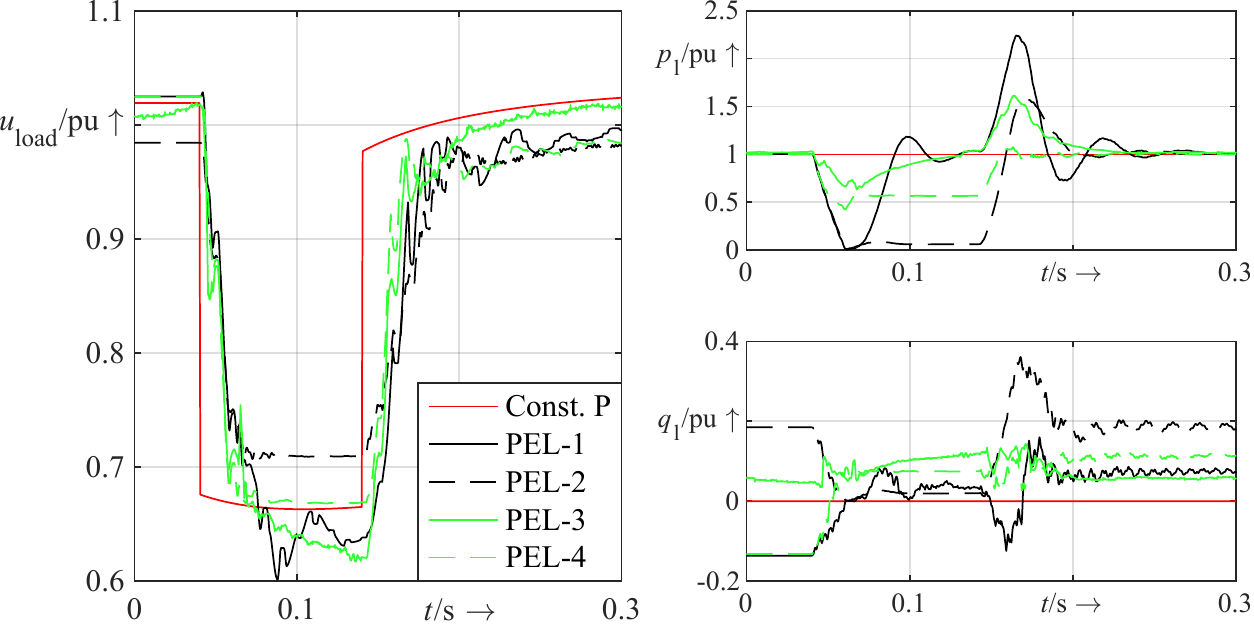}
    \captionsetup{justification=centering}
    \caption{Simulation results of scenario 1: $P_\mathrm{l} = 500\,$MW and $R_\mathrm{f} = 17\,\Omega$}
    \label{fig:stvs_1}
\end{figure}

Fig.\,\ref{fig:stvs_1} shows the first fault scenario, where the fault resistance is $R_\mathrm{f} = 17\,\Omega$ and $P_\mathrm{l} = 500\,$MW. 
As the results show, no \ac{PE} load or the constant power load leads to  instability. During the fault, the load voltage $u_\mathrm{load}$ of PEL-1 and PEL-3 is lower compared to the constant power load, while for PEL-4 it is slightly and for PEL-2 clearly higher. For PEL-1 and PEL-3 this can be explained by the fast power restoration during the fault and their inductive reactive power. Since the switch-off voltage from PEL-2 is relatively high, it is completely disconnected during this time, why no active and reactive power is consumed. However, before and after the fault, the voltage $u_\mathrm{load}$ is smaller compared to the other load, due to the high inductive reactive power consumption. This scenario demonstrates that some \ac{PE} loads lead to a higher and others to a lower voltage during the fault, compared to the constant power load. Here, the main reasons for a lower fault voltage are, if the load is capable of recovering its power during the fault and if its reactive power consumption is inductive. 

\begin{figure}
    \centering
    \includegraphics[width=1\columnwidth]{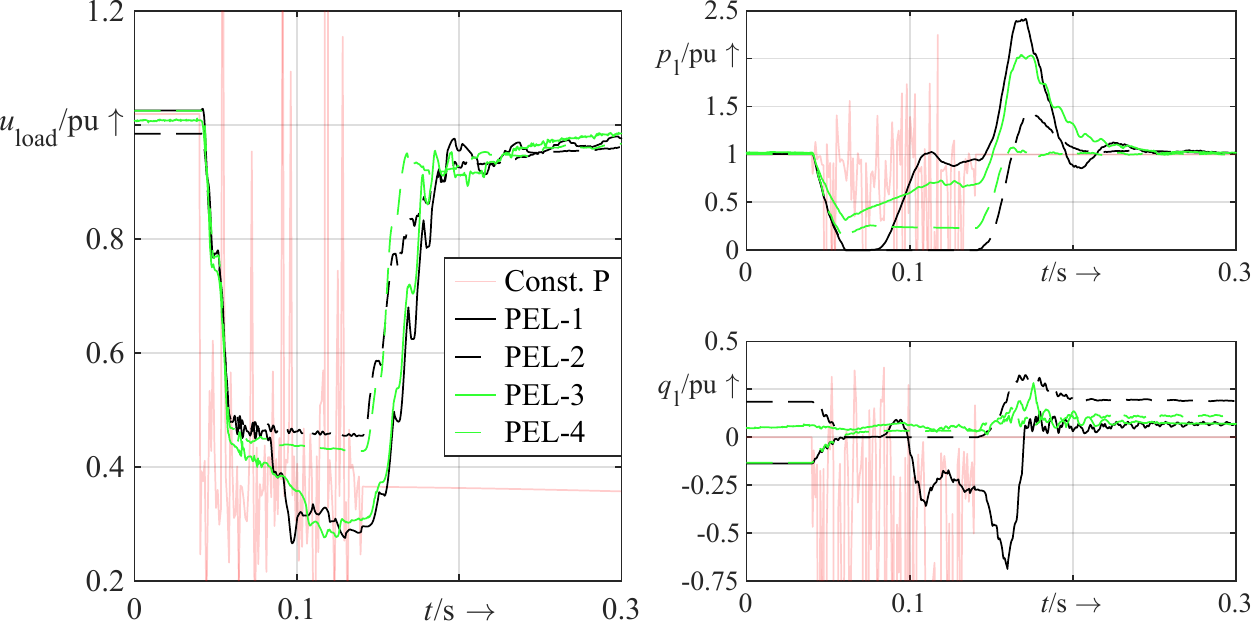}
    \captionsetup{justification=centering}
    \caption{Simulation results of scenario 2: $P_\mathrm{l} = 500\,$MW and $R_\mathrm{f} = 1\,\Omega$}
    \label{fig:stvs_2}
\end{figure}

In the second scenario, only the fault resistance is changed to $R_\mathrm{f} = 1\,\Omega$. Here it is investigated if a lower fault voltage leads to a higher power restoration at recovery. As can be seen by the results in Fig.\,\ref{fig:stvs_2}, the grid is not short-term stable for the constant power load, leading to numerical instability of the simulation. In contrast, for all \ac{PE} loads, the grid remains stable, as the decreasing power consumption helps to restore the voltage at the beginning of the fault.   
Yet, like in the simulations before, the voltage $u_\mathrm{load}$ for PEL-1 and PEL-3 is lower during the fault, compared to PEL-2 and PEL-4. The active and reactive power consumption after the fault is slightly higher for PEL-1 and PEL-3, compared to the previous simulation, but not for the other loads. Therefore, it can not be generally concluded that a smaller fault resistance or lower fault voltage, always leads to a higher power consumption at recovery.

\begin{figure}
    \centering
    \includegraphics[width=1\columnwidth]{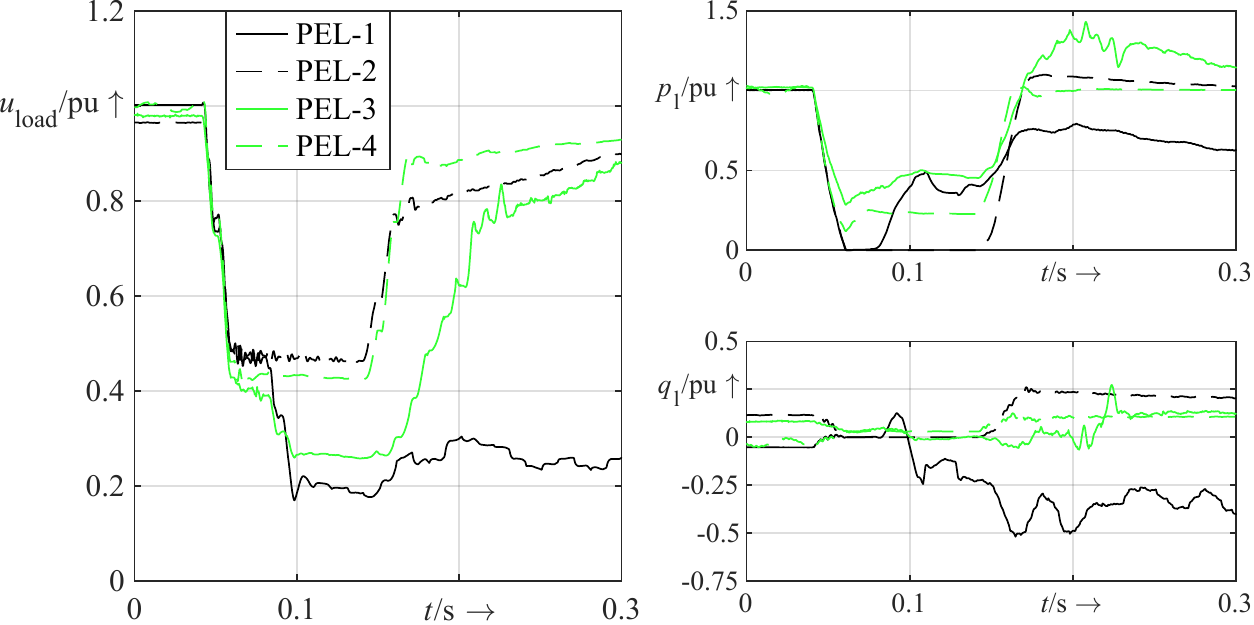}
    \captionsetup{justification=centering}
    \caption{Simulation results of scenario 3: $P_\mathrm{l} = 800\,$MW and $R_\mathrm{f} = 1\,\Omega$}
    \label{fig:stvs_3}
\end{figure}
For the last scenario, the rated active power of the \ac{PE} loads is increased to $P_\mathrm{l} = 800\,$MW, whereas the fault resistance is still $R_\mathrm{f} = 1\,\Omega$. Since the scenario with the constant power load was already unstable, this load is not simulated here. Fig.\,\ref{fig:stvs_3} shows the simulation results. Here, it can be seen that for PEL-1 the grid becomes unstable and can not restore its voltage after the fault, although its reactive power is capacitive. Also, for PEL-3 the voltage restoration is considerably delayed, due to its power restoration and its overshoot. In the case of PEL-2 and PEL-4, the voltage is similar to the previous simulation, but its recovery is more delayed. The similar voltage level can be explained by the disconnection of these loads. In conclusion, this scenario demonstrates that if the \ac{PE} is large and tries to recover its power during the fault, the grid can become unstable or voltage recovery can be considerably delayed. Moreover, the last two scenarios indicate that \ac{PE} loads have a significantly more positive influence on short-term voltage stability, compared to constant power loads.

\begin{table}[b]
\caption{Values of voltage stability test system (partly taken from \cite{Nordic})}
    \centering
    \begin{tabular}{c c c c c c}
    \hline
     & $S_\mathrm{base}$ & $U_\mathrm{base}$ & $X_\mathrm{l,line}$ & $R_\mathrm{line}$ &$X_\mathrm{c,line} / 2 $ \\
    Grid/lines & 100 MW & 380 kV  & $64\,\Omega$ & $9.6\,\Omega$ & $1334.63\,\Omega$\\  
    \hline
    &  $S_\mathrm{r,T1}$ & $u_\mathrm{k,T1}$ & $R/X_\mathrm{T1}$ & $n_\mathrm{tap,T1}$ & $\Delta u_\mathrm{tap}$ \\
    T1 &1.2 GVA & 0.15 pu & 0.0 & 1 & 0.01\,pu\\
    \hline
    & $S_\mathrm{r,T2}$ & $u_\mathrm{k,T2}$ & $R/X_\mathrm{T2}$ & $n_\mathrm{tap,T1}$ & $\Delta u_\mathrm{tap}$\\
   T2 & 600 MVA & 0.15 pu & 0.0 & 0 & 0.0\,pu \\
    \hline
    \end{tabular}
    
    \label{tab:3bus}
\end{table}

\section{Conclusion}
In this paper, comprehensive research about the fundamental power response and dynamic modelling aspects of \ac{PE} loads in case of voltage drops has been conducted. For this, the power response of four loads has been measured in the laboratory. A focus is set on \ac{SMPS} loads with different \ac{PFC} techniques. The results show that the power response depends on the used \ac{PFC} technique and the intensity of the voltage drop. In contrast, the fault duration has only a small impact, as the \ac{PE} loads reach an intermediate steady-state during the fault. In the laboratory, also the influence of the grid impedance and \ac{EMI} filter on the power response have been measured. Here, it could be shown that a high grid impedance has a strong impact, resulting in delayed power recovery and earlier disconnections from the grid. In the case of the \ac{EMI} filter, it is demonstrated that it has nearly no influence on the fundamental active power and only a small, but negligible impact on fundamental reactive power. The results also highlight that \ac{PE} loads should be simulated in the EMT domain, as the dynamics of the inductances and capacitances of the grid and \ac{PE} loads have to be considered. 
 
Based on the measurements, simulation models for each \ac{PE} load have been derived. In addition, a simple switch-off control is introduced, which simulates the disconnection of the loads in case of too low voltages. With the help of scientific literature, it is also pointed out according to which aspects the components of the respective loads are dimensioned. This allows the simulation models to be parameterised even without explicit knowledge about specific loads.
In the next step, the simulation models are manually tuned to meet the measurements. In general, the simulation models capture the main dynamics of the loads, but with some differences. On the one hand, this can be explained by the limitation of the laboratory equipment. On the other hand, no methodical optimisation of the parameter tuning has been carried out. Nevertheless, the accuracy achieved here is sufficient and shows which aspects are important when modelling \ac{PE} loads for voltage drops. 
At the end, a use case of the developed models is shown, where their influence on short-term voltage stability is investigated and their differences to a constant power load are highlighted. The results indicate that \ac{PE} loads have a more positive impact on short-term voltage stability, as they reduce their power consumption during the fault. However, if the \ac{PE} load is big enough, voltage instability is still possible due to its power recovery. In addition, voltage recovery can also be significantly delayed which could also arise other instabilities. Generally, the switch-off voltage parameter of the loads has a great impact on voltage stability, as it determines when and if power recovery takes place during the fault. 

In the future, other \ac{PE} loads like LED or electric vehicle charges will be measured to extend the modelling aspects. Also, additional \ac{SMPS} loads shall be measured to identify, if common load parameters can be derived. In addition, measurements of similar loads could also confirm that the chosen load structures are suitable. 
As the voltage stability investigation in this paper is limited, more complex grids shall be taken into account to further analyse the impact of \ac{PE} loads. Here the interplay between \ac{PE} loads and induction machines is of special interest, due to the fault induced delayed voltage recovery phenomena. Moreover, the share of \ac{PE} loads of the total load should be quantified and how big the share of \ac{SMPS} loads is. 
As the paper mainly deals with the fundamental power for voltage stability, these loads may have to be modelled differently for other stability types, like harmonic stability. 
\section*{Acknowledgment}
This work was funded by the Deutsche Forschungsgemeinschaft (DFG, German Research Foundation) - 360460668.

\bibliography{lib.bib}
\bibliographystyle{IEEEtran}

\end{document}